Black Holes and Vacuum Cleaners: Using Metaphor, Relevance, and Inquiry in

Labels for Space Images

Lisa F. Smith

University of Otago

Kimberly Kowal Arcand

Chandra X-ray Center/Smithsonian Astrophysical Observatory

Benjamin K. Smith

University of Mary Washington

Randall K. Smith

Chandra X-ray Center/Smithsonian Astrophysical Observatory

Jay Bookbinder

NASA Ames Research Center

Jeffrey K. Smith

University of Otago

Corresponding Author:
Professor Lisa F. Smith University of Otago College of Education
145 Union Street East
Dunedin 9054 New Zealand
+64 3 479 9014
professor.lisa.smith@gmail.com

**Abstract**

This study extended research on the development of explanatory labels for astronomical images for the non-expert lay public. The research questions addressed how labels with leading questions/metaphors and relevance to everyday life affect comprehension of the intended message for deep space images, the desire to learn more, and the aesthetic appreciation of images. Participants were a convenience sample of 1,921 respondents solicited from a variety of websites and through social media who completed an online survey that used four high-resolution images as stimuli: Sagittarius A*, Solar Flare, Cassiopeia A, and the Pinwheel Galaxy (M101). Participants were randomly assigned initially to 1 of 3 label conditions: the standard label originally written for the image, a label with a leading question containing a metaphor related to the information for the image, or a label that contained a fact about the image relevant to everyday life. Participants were randomly assigned to 1 image and compared all labels for that image. Open-ended items at various points asked participants to pose questions to a hypothetical astronomer. Main findings were that the relevance condition was significantly more likely to increase wanting to learn more; the original label was most likely to increase overall appreciation; and, smart phone users were more likely to want to learn more and report increased levels of appreciation. Results are discussed in terms of the need to examine individual viewer characteristics and goals in creating different labels for different audiences.

Black Holes and Vacuum Cleaners: Using Metaphor, Relevance, and

Inquiry In Labels for Space Images

In going about our everyday lives, we encounter beauty in a variety of forms. It might be in a sunset, a bird of brilliant color, or a picturesque Italian village. But within the past 50 years, it might also be in an astronomical image that while stunning, is also difficult to comprehend. Such images have become part of everyday life for millions of people around the world. The Astronomy Picture of the Day (APOD) website gets over 1 million visitors a day to their website. The Chandra telescope gets about 9-10 million visitors a month, and NASA has 20 million followers on Twitter (Arcand, personal communication). Add to these numbers the 5.8 million weekly visitors of Cosmos: A Spacetime Odyssey on television, and too many Star Wars/Star Trek aficionados to count. Astronomical imagery is ubiquitous.

But it isn't always particularly comprehensible. Consider Cassiopeia A (Cas A). Cas A is a supernova remnant and images of it are spectacular (see Figure 1). Scientists and members of the lay public alike look on images of Cas A with awe, wonder, and appreciation for its aesthetic appeal. But, beyond the initial "wow" factor, what is really stunning about supernova remnants like Cas A is that they were once massive stars that ran out of fuel, collapsed, exploded, and propelled their remains into space. Those remains contained elements such as oxygen and iron that not only helped to form new stars and planets, but eventually created the oxygen we breathe, the iron in our blood, and the calcium in our bones. In fact, those nuclear "furnaces" are the only place that scientists know of where such vital elements for life can be made. We are, indeed, the stuff of stars.

Once we learn that life as we know it basically came from stars and their violent deaths, the "wow" becomes "WOW!" That information, which links the person viewing the image to cosmic events from the distant – or not so distant - past, draws us into the image and makes the astronomical, personal. There are many interesting pieces of information that might be communicated to the public about Cas A. It is roughly 11,000 light years away, and the explosion of Cas A occurred only about 300 years ago. The part that is visible in the image is roughly 50 million degrees hot and what we see is expanding at about 5000 km/second. This is not a "still" picture. And yet, it seems that the real "grabber" for this image is probably the fact that parts of us were once parts of that (or, more precisely, a similar "that"). The challenge that faces professionals in charge of public communications and engagement for the space program is how to convey scientific information in an engaging, educational, clear, concise, and non-condescending way.

Astronomical images might be considered the essence of taking scientific data and conveying it as aesthetically pleasing visual representations. Although some have considered images from space as just another "pretty face(s)," (Cendes, 2007, para. 2), others have argued for the actual scientific value that impart the "scientist's objectives" (Robin, 1992, p. 9). Though the role of the image is a complex one (Trumbo, 2000), it is recognized as a fundamental part of communicating science in general (Frankel, 2004; Nicholson-Cole, 2005), and astronomy in particular (Kessler, 2012). Images can, however, "overpower words" (Lazard & Atkinson, 2014 p. 11); nowhere is this truer than in astronomical images, so communicating with intent becomes important. This brings us to the issue of how to achieve that communication in combination with the aesthetics, especially with those who may not have expertise in astronomy.

Many choices must be made in the creation of the image itself from data, factoring in color choices and sharpening of the image, and creating the context for that image and its description (Arcand et al., 2013, Rector, Levay, Frattare, English, & Pu'uohau-Pummill, 2007). Astronomical images are not "records of the real," (Rothstein, 2010, para. 11) but the result of a series of such choices - from the field of view selected, to the energy cuts of the data that are included in producing the image, to rendering a careful balance between scientific and aesthetic viewpoints (Arcand et al., 2013; Kessler, 2012; Rector, Arcand, & Watzke, 2015). Thus, the astronomical image is something of a hybrid: in some respects a graph, and in others, a work of art. It is the presentation of data, sometimes multiple sets of data, in such a way that it can be analyzed by scientists and at times awe-inducing to the wider community.

For such data to be accessible for many different audiences, it must be informative, understandable, and relatable. Communicating astronomy with non-experts quickly gets complicated, especially once we move past our familiar solar system and other "nearby" objects that we can see in our night sky (Smith et al., 2011). Scales move out of the realm of the lay public's reference points, quickly going from millions of kilometers to millions of light years. When the mass of our own star, the Sun (a somewhat small and dull star in relation to the rest of the stars in the Universe) weighs in at two nonillion kilograms, we need to come up with ways to make that information comprehensible.

People approach such scientific data from their existing frames of reference (Wynne, 1991), with research strongly suggesting that knowledge and reasoning ability in mathematics and science form a context or lens through which they understand images (e.g., Brown, Collins, & Duguid, 1989; Carraher, Carraher, & Schliemann, 1985; Lave, 1988; Osborne, 2007). Indeed, Smith and Wolf (1996) found that three things are influential to viewers' experiences and level of engagement in a museum: the exhibit objects, the presentation of those objects, and the prior knowledge of the viewer(s). More recently, Lazard and Atkinson (2014) advised that layers of expertise, experience, and understanding should be considered when using visuals to communicate a particular message. Estrada and Davis (2015) stated that the role of the texts in describing the data visualization is key, finding that "All texts, including visual elements, are consciously constructed and have particular social, cultural, political and economic purposes" (p. 142). Freise (2016) also noted that contextualization - recognizing people's values, experiences, and beliefs – is critical for science communication.

Although research on labels in museums has been conducted for decades (see Cogswell, 2015 for a recent overview), especially in terms of how to write labels (e.g., Bitgood, 2013; Falk & Dierking, 2013; Serrell, 2015), accessibility for particular groups such as those with low vision (Wolf & Smith, 1993), and for specific types of museums (e.g., Borun & Miller, 1980; Cogswell, 2015), there is scant experimental research on the content of labels in general, and labels that accompany astronomy images in particular.

Given that how the communication is presented to viewers is important,

the question then turns to what the optimal options are, which in turn engenders the question of "optimal for what?" We turn to work that has been conducted in other areas for help, in particular, to communication concerning works of art. Although one might argue that there are important differences between works of art and images from astronomy, it seems a reasonable place to look for insight. Research on how accompanying information influences perception and reception of artworks has a fairly long history (see Kreitler & Kreitler, 1972). As Belfiore and Bennett (2007) argued, there are a host of factors that influence the encounter of an individual with a work of art, and the label and title of the work are among those.

    Russell (2003) looked at the effects of providing titles and brief descriptions of works of art on their perceived meaningfulness and pleasingness. An effect on meaningfulness was found in both studies reported, but on only one of the studies for pleasingness. Leder, Carbon, and Ripsas (2006) provided a very useful distinction between labels that are descriptive and those that are elaborative. They defined elaborative as those labels that provided "an explanation or a metaphoric interpretation of the scene" (p. 178), and descriptive as labels that simply provided a title representing what was in the image. They found that when provided enough time to process the image the way it would typically be done in a museum (Smith & Smith, 2001), elaborative titles led to greater understanding of the works.

    Looking at the effects of labels on visitor behavior has long been studied within the field of museology. Hirschi and Screven (1988) pioneered a series of studies of the motivating effects of the use of leading questions in museum labels. Jones (1995) found that using questions increased

engagement, and Litwak (1996) concluded that labels with leading questions were motivating for visitors to the Bell Museum of Natural History. Gutwill (2006) found that visitors who were interviewed at The Exploratorium seemed to prefer a combination of questions and suggestions for labels that accompanied an interactive exhibit.

Since 2008, the Aesthetics and Astronomy group (A&A) has investigated different types of label content used with astronomical images. A&A has explored public perceptions of astronomical images and how best to communicate their underlying science to the public. A&A is made up of astrophysicists, image development professionals, and educators with expertise in research methodology. In a series of studies (Smith, 2014; Smith et al., 2011; Smith et al., 2014; Smith, Smith, Arcand, Smith, & Bookbinder, 2015), a variety of label approaches has been examined in experimental and non-experimental settings. Support consistently has been found for the leading question format, along with labels that presented some interesting or amusing "tie-in" from the astronomical image to everyday life, and those that employed metaphors related to everyday life. This study extends that work, along with what has been learned from other research, to propose powerful approaches that will maximize interest in, and comprehension and aesthetic appreciation of astronomical images.

To that end, in this study we contrasted three label types. First, we employed the standard labels generated to accompany images as they were released to the public. This might be analogous to what Leder et al. (2006) referred to as *descriptive titles*. Then, we developed two approaches to alternative forms of labels, those that the Leder group might call *elaborative* *titles*. In the first elaborative approach, we combined the leading question idea with the use of metaphor. Metaphor can be particularly powerful in trying to convey ideas in physics that can be hard for the lay person to grasp (Zides, 2015). In the second approach, we refined our previous work of trying to find a "tie-in" and focused on developing a label that held the potential for being personally relevant to the viewer. We sought to "bridge the gap" (Nicholson-Cole, 2005, p. 255) between what might be the abstract or difficult to comprehend within astronomical images and the everyday lives of viewers of those images.

Therefore, our objective for this research was to examine the effects of three approaches to writing labels for astronomical images in terms of comprehension, engagement, and aesthetic appreciation, taking into consideration what viewers might still want to know. As such, our research questions were:

1. How does the use of leading questions/metaphors and relevance to everyday life affect the comprehension of astronomical images?
2. How do leading questions/metaphors and relevance to everyday life in labels affect levels of engagement assessed by participants wanting to learn more for astronomical images?
3. How do leading questions/metaphors and relevance to everyday life in labels affect aesthetic appreciation for astronomical images?
4. What kinds of questions do viewers have after receiving different types of labels accompanying the images?

## Method

### Participants

Participants were a convenience sample of 1,921 respondents to an online survey. Using an open-ended format for gender, there were *n* = 1,018 (53%) males, *n* = 615 (32%) females, *n* = 9 (.5%) queer, *n* = 1 (.1) each non-binary and transgender; *n* = 277 (14.4%) did not respond to this item. Age was roughly evenly distributed with *n* = 323 (16.8%) for 18-24 years, *n* = 250 (13%) for 25-34 years, *n* = 267 (13.9%) for 35-44 years, *n* = 245 (12.8%) for 45-54 years, *n* = 335 (17.4%) for 55-64 years, and *n* = 294 (15.3%) for 65 and above, with *n* = 207 (10.8%) not responding.

Participants were well-educated, with *n* = 242 (12.7%) reporting having earned an advanced degree (law, medical or doctorate), *n* = 391 (20.4%) a postgraduate or masters degree, and *n* = 512 (26.7%) an undergraduate degree. The remaining participants reported their highest level of education as having some university (*n* = 331; 17.2%), a secondary/high school degree (*n* = 188; 9.8%), and some secondary/high school (*n* = 49; 2.6%). There were *n* = 208 (10.8%) non-respondents to this item.

A variety of occupations were reported. The most frequently reported were retired (*n* = 249; 13.0%), student (*n* = 206; 10.7%), computer/technical (*n* = 197; 10.3%), education/librarian (*n* = 154; 8.0%), arts/entertainment/sports (*n* =105; 5.5%), medical/health (*n* = 90; 4.7%), astrophysicist/astronomy-related (*n* = 88; 4.6%), government (*n* = 54; 2.8%), finance (*n* = 52; 2.7%), trades/construction (*n* = 42; 2.2% each), service-related field (*n* = 42; 2.2% each), and executive/management (*n*= 42; 2.2%).

Participants were asked about their knowledge of and background in astronomy. In terms of self-reported knowledge of astronomy, using a scale from 1 (complete novice) to 10 (expert), the mean response was 5.26 ($SD =$ 2.35). Responses were normally distributed, with fewer experts than novices at the extremes. For background in astronomy, participants could give multiple responses. A total of $n = 1,175$ (61.2%) participants reported that they read about astronomy online; $n = 833$ (46.0%) reported that it was a hobby; $n = 448$ (23.3%) reported having studied astronomy in secondary/high school; $n = 422$ (22.0%) had taken one or more university courses in astronomy; $n = 159$ (8.3%) participated in an amateur astronomy organization; $n = 122$ (6.4%) held a degree in astronomy or worked in a related field; and $n = 388$ (20.2%) reported having no background.

Given the online nature of the survey, we were interested in what type of device the participants were using to view the images. Of the $n = 1,721$ who completed this item, $n = 588$ (30.6%) used a desktop computer, $n = 510$ (26.5%) used a smartphone, $n = 500$ (26.0%) used a laptop computer, $n = 88$ (4.6%) used a full-sized tablet, and $n = 35$ (1.8%) used a mini-tablet.

We were also interested in how participants learned about the survey. In an effort to get a diverse sample of participants, we solicited participation through sites where individuals interested in astronomy were likely to go, but also to sites unrelated to astronomy. Of the $n = 1,720$ responses, the most popular response was the Astronomy Picture of the Day website ($n = 811$; 42.2%), followed by the Chandra website/Chandra Instagram ($n = 226$; 11.8%). Other social media, not specified by the respondent was next ($n = 205$; 10.7%), followed by word of mouth ($n = 147$;

7.7%), Facebook (*n* = 129; 6.7%), Twitter (*n* = 78; 4.1%), and Smithsonian email (*n* = 74; 3.9%). The remaining responses were comprised of the Aesthetics & Astronomy website, Craigslist, other news outlet, and other-not specified (*n* = 50; 2.6%).

**Materials**

A survey was written for this study using four high-resolution images as stimuli: Sagittarius A*, Solar Flare, Cassiopeia A, and Pinwheel Galaxy (M101), as shown in Figure 1. These were chosen to represent distinct types of images. There were three labels for each image. One was the standard label that had been originally written for the image. The other two labels were written by the researchers, with an effort made to have roughly the same word count as the standard label. One approach included a leading question and a metaphor related to the information for the image. The other approach had information presented in the context of a fact that was relevant to individuals in everyday life. All labels are shown in the Appendix. The items on the survey are explained in the Procedure section, as it facilitates reading and understanding the design.

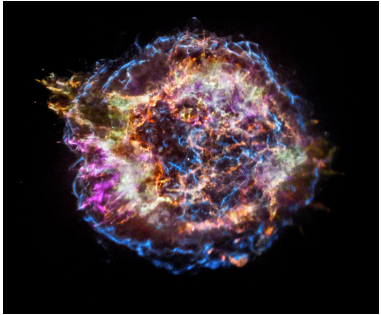

Cassiopeia A
Credit: NASA/CXC/SAO

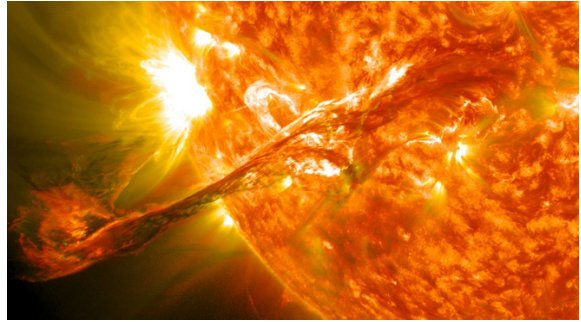

Solar Flare Credit: NASA/GSFC/SDO

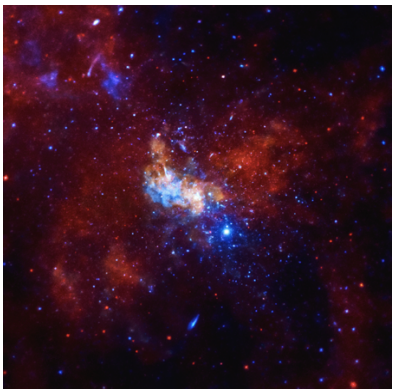

Sagittarius A*
Credit: NASA/CXC/Univ. of Wisconsin/Y.Bai et al.

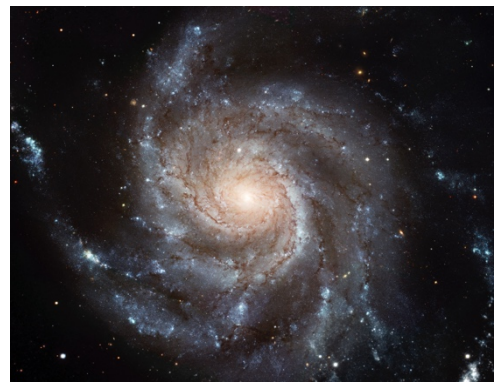

Pinwheel Galaxy (M101)
Credit: NASA, ESA, K. Kuntz (JHU), F. Bresolin (Univ. Hawaii), J. Trauger (JPL), J. Mould (NOAO),
Y.-H. Chu (Univ. Illinois, Urbana), STScI

*Figure 1*. Images used in the survey.

**Procedure**

Participants were solicited from a variety of websites chosen to represent those with an interest in astronomy as well as the lay public. Websites included the Chandra X-ray Observatory (chandra.si.edu); the Astronomy Picture of the Day (apod.nasa.gov); the Aesthetics and Astronomy website (http://astroart.cfa.harvard.edu/); Craigslist; the Boston Globe; social media sites such as Facebook, Twitter, and Instagram; and, museum networks known to the researchers, such as the American

Psychology Association Division 10 and the International Association of Empirical Aesthetics listservs.

If individuals reading the invitation to participate on one of those websites chose to participate, they clicked a button that took them to a site developed to administer the survey. On the first page of the survey, participants were advised that those age 18 or over were invited to continue and that submission of the survey indicated consent. This study was deemed exempt by the Human Subjects Institutional Review Board of the Smithsonian Institution.

The design of the study is somewhat complicated, and therefore is presented in schematic form in Figure 2. The survey began with participants being randomly assigned to two groups, A and B. Group A responded to a set of six knowledge items *prior* to looking at the images and accompanying labels. These items asked participants to rate on a scale of 1 (nothing) to 10 (expert knowledge) how much they knew about the objects to be depicted in the four images in the study plus two objects that were not part of the study: hyperplanets, and Wolter nebulae. Neither hyperplanets nor Wolter nebulae are actual objects; they were made up for the study. Group B responded to the same set of items at the end of the survey, after they had seen the images and read the labels. The idea of using self-ratings in a pre/post fashion, along with objects not in the study, has been used successfully previously in research on art perception (Smith & Smith, 2003).

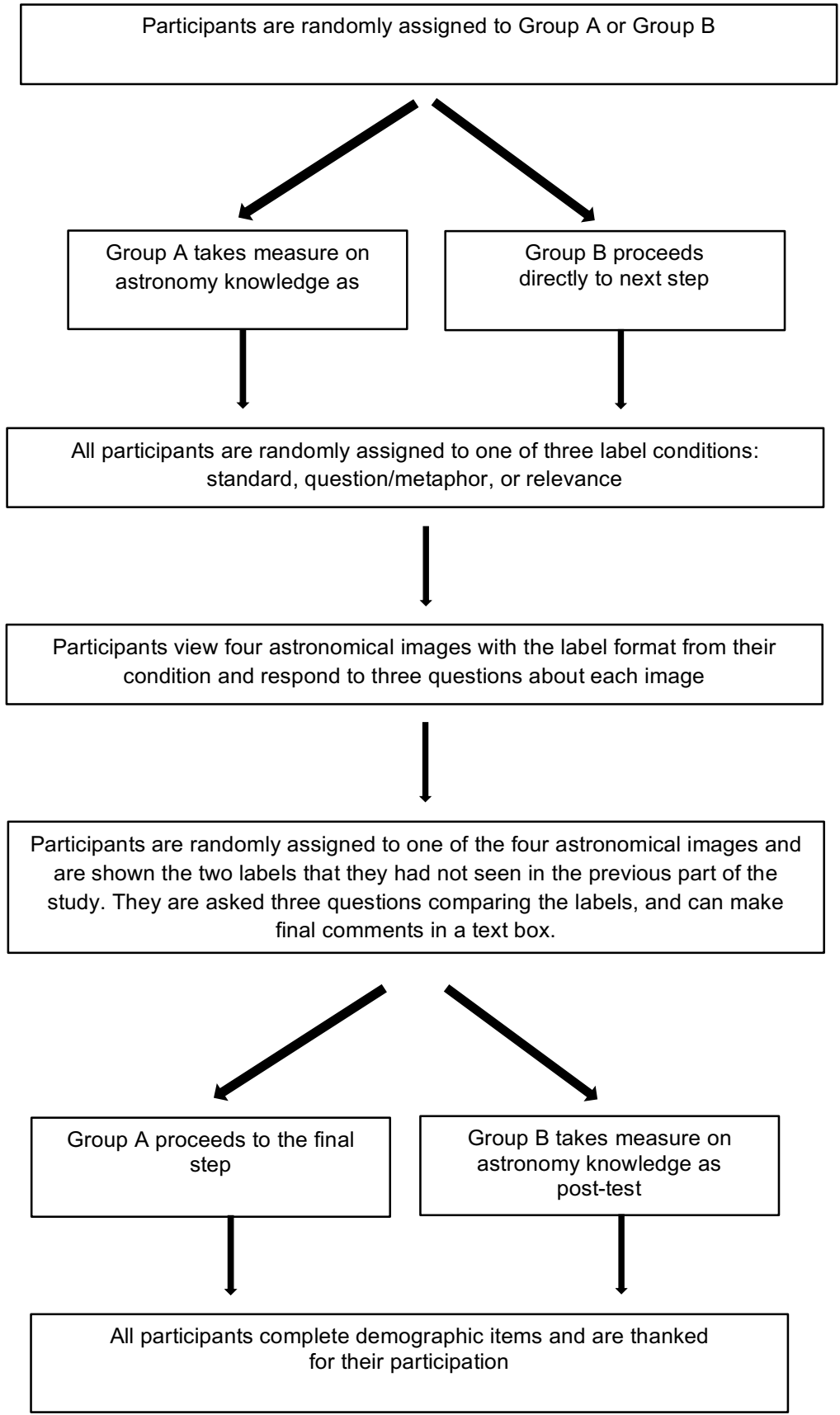

*Figure 2.* Schematic of study design.

All participants were then randomly assigned to one of the three types of labels (described above). They viewed all four images with their assigned label type. After each image and its assigned label, participants were asked three questions:

1. Using a scale of 1 (not at all) to 10 (definitely), how much do you want to learn more about this topic?

2. Using a scale of 1 (not at all) to 10 (a great deal), how much does this explanation increase your appreciation of the image?

3. What question might you ask if an astronomer were with you now? (Responses to this question were recorded in a text box.)

Next, participants were randomly assigned to view *one* of the four images, with the two types of labels for that image that were not previously assigned, along with the label that they had already evaluated. They were instructed:

"Here are two other explanations for [name of image], along with the explanation you've already seen. Thinking about all three explanations, please respond to these questions:

1. Which provides the best explanation for the image? [label 1, label 2, or label 3]

2. Which explanation makes you want to learn more about this topic? [label 1, label 2, or label 3]

3. Which explanation increases your overall appreciation of the image best? [label 1, label 2, or label 3]"

After completing these questions, Group B took the six-item knowledge measure that Group A had taken at the beginning of the survey.

This allowed for looking at pre/post differences as a result of having viewed the images and read the labels. All participants responded to seven demographic items that asked for a self-rating of knowledge of astronomy, background in astronomy, age group, highest level of education, gender, occupation, type of computer platform used for the survey, and how they learned about the survey.

A final open-ended item invited participants to describe anything they wanted to add about the survey.

The survey took approximately 15 minutes to complete.

The quantitative data were analyzed using SPSS Version 22. For the qualitative data, there were 3 sets (label type) by 4 (images) of qualitative responses obtained, and a set of responses for the final open-ended item. Using Strauss and Corbin's (1998) grounded theory approach, two researchers independently developed a series of codes, and compared and agreed upon a final set of codes for each image. Each response was then independently rated by the two researchers and checked for inter-rater reliability. There were few discrepancies, which were resolved by discussion between the raters.

## Results

**Label Type and the Comprehension of the Intended Message**

The first research question asked how the use of leading questions/metaphors and relevance to everyday life affect the comprehension of the intended message in labels for deep space images. Participants were asked to rate their knowledge of six astronomical objects on a 1 (nothing) to 10 (expert knowledge) scale; four of these were the images presented in the study, and two were not actual objects, but made up for this study by an astrophysicist (one of the authors) to sound like real astronomical objects. We summed the ratings on the four real objects to make a scale. The scale had a coefficient alpha reliability for the sample of .91. We then summed the two objects that were not real to make a second scale; this had a coefficient alpha of .71 for these participants.

These two knowledge scales were then used as dependent variables in two three-way analyses of variance (ANOVA). The independent variables were pre/post administration of the questionnaire (2 levels), type of label read (3 levels), and level of expertise (3 levels). For level of expertise, the response to the question, "How much do you know about astronomy?" (1 = Novice to 10 = Expert) was made into three categories (Low = 1-3, 27.2%; Medium = 4-6, 39.1%; and High = 7-10, 33.7%). The first ANOVA was run on the knowledge scale that was based on the four objects that were used in the study. Means and standard deviations are presented in Table 1.

This analysis was significant, $F(17, 1703) = 145.55$, $p < .001$. The interaction of interest was the two-way interaction between pre/post administration and type of label. This interaction was significant, $F(2, 1703) = 4.22$, $p = .015$. The main effect for expertise was also significant, $F(2, 1703) = 1226.59$, $p < .01$. This was expected, as it merely showed that people with more expertise reported that they knew more about the objects. The main effect for pre/post was also significant, $F(1, 1703) = 4.17$, $p = .041$, with the post means slightly higher than the pre means.

Black Holes and Vacuum Cleaners 19Table 1

*Descriptive Statistics for ANOVA on the Four Item Knowledge Scale*

| Type of Label | Expertise Group | Pre | | | Post | | |
|---|---|---|---|---|---|---|---|
| | | n | M | SD | n | M | SD |
| QA/Metaphor | Low (1-3) | 83 | 3.07 | 1.41 | 71 | 2.84 | 1.02 |
| | Medium (4-6) | 106 | 4.68 | 1.33 | 108 | 4.69 | 1.13 |
| | High (7-10) | 96 | 6.85 | 1.28 | 95 | 6.97 | 1.45 |
| Relevance | Low (1-3) | 75 | 2.71 | 1.15 | 72 | 3.13 | 1.24 |
| | Medium (4-6) | 108 | 4.72 | 1.38 | 120 | 5.15 | 1.29 |
| | High (7-10) | 105 | 6.76 | 1.28 | 88 | 7.07 | 1.25 |
| Standard | Low (1-3) | 75 | 2.93 | 1.10 | 84 | 2.87 | 1.25 |
| | Medium (4-6) | 138 | 4.87 | 1.39 | 100 | 4.77 | 1.28 |
| | High (7-10) | 101 | 6.66 | 1.36 | 96 | 6.92 | 1.23 |

Looking at the pre/post mean differences for the three label types, the relevance label led to the greatest difference, with an effect size (Cohen's *d*) of .299. The effect size for the pre/post difference for the question/metaphor label was .080, and for the standard label was -.081. The interaction of pre/post with type of label is depicted in Figure 3.

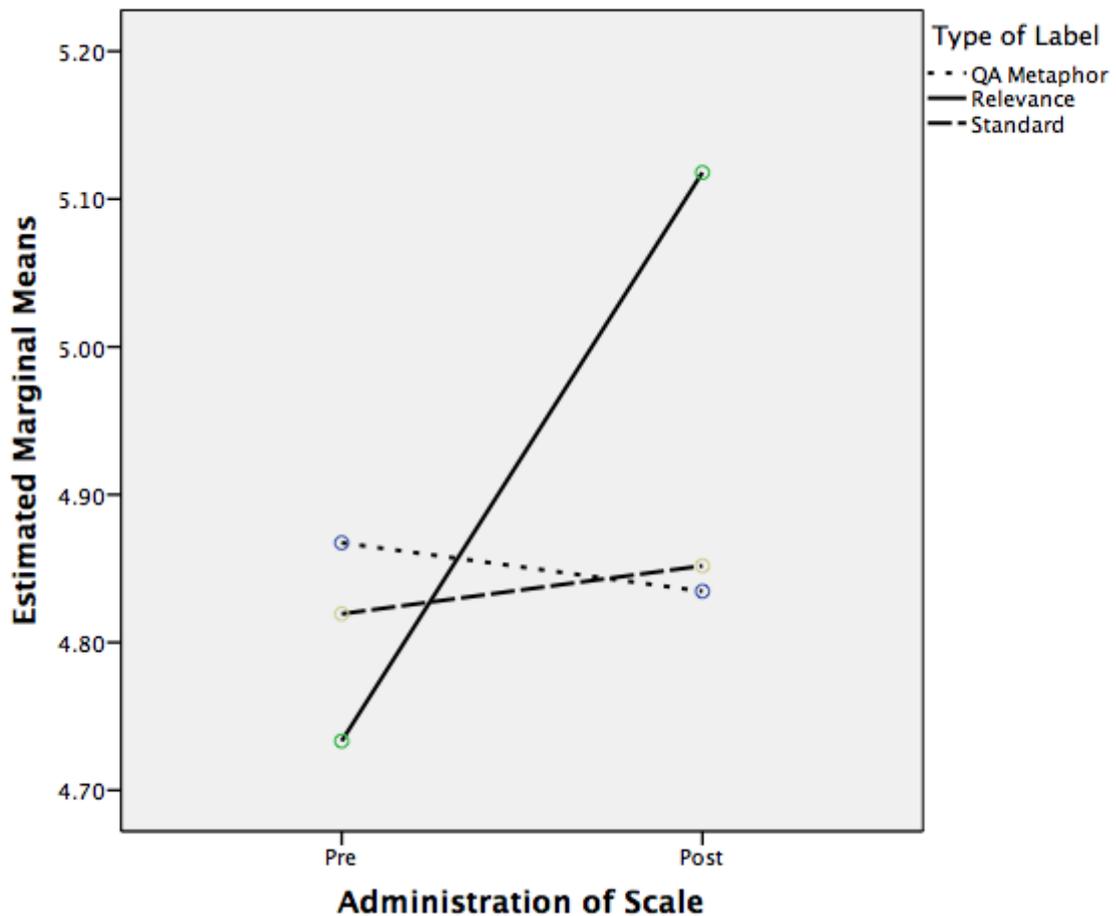

*Figure 3.* Interaction of pre/post administration of four item knowledge measure with type of label.

A similar analysis was run with the two objects that participants rated that were made up for this study (not actual astronomical objects). Here, no growth from pre to post was expected. Although the total model showed statistically significant results, $F(17, 1706) = 21.59$, $p < .001$, only the main effect for expertise was significant, $F(2, 1706) = 176.92$, $p < .001$. Interestingly, participants with higher levels of self-reported expertise reporting knowing more about the non-existent objects than those with less expertise.

These analyses might have been conducted using self-reported expertise as a covariate rather than breaking it into three groups and entering those groups as a separate factor. We ran the analyses both ways with highly similar results. We therefore chose the "factor approach," as that allowed us to look more easily at non-linearity and at interactions with other variables.

**Categories Used for Subsequent Analyses**

Prior to analyzing the data for the remaining research questions, it was necessary to collapse the categories for gender and type of computer platform. For gender, only those participants who identified as male or female were retained. For type of computer platform, mini-tablet and full-sized tablet were combined into one category. This was based on independent samples *t*-tests using mini-tablet and full-sized tablet as the independent variables, and age group, level of expertise, gender, and the measures for wanting to learn more and appreciation as dependent variables, all of which yielded non-significant differences. Age was collapsed into three groups of 18-34 ($n = 540$, 33.1%), 35-54 ($n = 490$, 30.0%), and 55 and above ($n = 597$, 36.6%).

**Label Type and Wanting to Learn More**

The second research question explored how labels with leading questions/metaphors and relevance to everyday life affect the desire to learn more about deep space images. To examine this question, a 3 (label type) * 2 (gender) * 3 (age groups) * 3 (expertise) * 4 (computer platforms) analysis of variance (ANOVA) was used with responses to the item, "Using a scale of 1 (not at all) to 10 (definitely), how much do you want to learn more about this topic?" as the dependent variable (summed across for four images). The responses to the four individual items regarding wanting to learn more item were checked for reliability; alpha = .84 was obtained. As such, it was determined that there was sufficient internal consistency in responses across the four images to average responses for this analysis. Due to the size of the sample and the number of variables in the analysis, the alpha level was set to .01.

In the full factorial analysis, main effects were obtained for type of label, expertise, and computer platform; there were no main effects for gender or age, and no interaction effects. Given those results, the analysis was run again with only main effects in the model. Results are shown in Table 2. A Scheffé post-hoc procedure (alpha = .01) was used to examine the levels of the main effects. For type of label, those who were in the relevance condition were significantly more likely to want to learn more as compared to QA/metaphor. The effect size for this difference (Cohen's *d*) was .263. No other comparisons were significant. For expertise, those who reported the lowest levels of expertise (rating of 1-3) were less likely to want to learn more as compared to those who rated their expertise as 4-6 or 7-10. No other comparisons were significant. For type of computer platform, those using a smart phone were more likely to want to learn more as compared to the other types of platforms. The effect sizes were 1.94 (compared to tablet), .306 (compared to desktop), and .376 (compared to laptop). No other comparisons were significant.

Table 2

*ANOVA Main Effects for 3 (label type) * 2 (gender) * 3 (age groups) * 3 (expertise) * 4 (computer platforms) for Wanting to Learn More*

| Source* | n | M | SD | df | F | $\eta^2_p$ | p |
|---|---|---|---|---|---|---|---|
| Label Type | 1,626 | | | 2 | 6.11 | .008 | .002 |
|     QA/Metaphor | 529 | 7.54 | 1.77 | | | | |
|     Relevance | 536 | 7.91 | 1.63 | | | | |
|     Standard | 561 | 7.73 | 1.73 | | | | |
| Gender | 1,626 | | | 1 | 4.77 | | ns |
|     Male | 1,015 | 7.69 | 1.71 | | | | |
|     Female | 611 | 7.81 | 1.75 | | | | |
| Age Groups | 1,626 | | | 2 | 3.84 | | ns |
|     18-34 | 540 | 8.05 | 1.68 | | | | |
|     35-54 | 490 | 7.61 | 1.72 | | | | |
|     55 and above | 596 | 7.54 | 1.73 | | | | |
| Grouped Expertise | 1,626 | | | 2 | 21.01 | .025 | <.001 |
|     1-3 | 443 | 7.37 | 1.97 | | | | |
|     4-6 | 637 | 7.79 | 1.62 | | | | |
|     7-10 | 546 | 7.96 | 1.58 | | | | |
| Computer Platform | 1,626 | | | 3 | 9.91 | .018 | <.001 |
|     Smart Phone | 485 | 8.12 | 1.66 | | | | |
|     Tablet | 115 | 7.79 | 1.72 | | | | |
|     Laptop | 474 | 7.48 | 1.77 | | | | |
|     Desktop | 552 | 7.60 | 1.69 | | | | |

*NOTE.* No interaction effects were obtained.

**Label Type and Aesthetics**

    The third research question explored how labels with leading questions/metaphors and relevance to everyday life affect aesthetic appreciation for deep space images. To examine this question, a 3 (label type) * 2 (gender) * 3 (age groups) * 3 (expertise) * 4 (computer platforms) analysis

of variance was used with responses to the item, "How much does this explanation increase your appreciation of this image?" as the dependent variable. As with learning, the responses to the four individual appreciation items (summed across images) were checked for reliability; an alpha = .83 was obtained for these data. As such, it was determined that there was sufficient consistency in responses across the four images to average responses for this analysis. Due to the size of the sample and the number of variables in the analysis, the alpha level was set to .01.

In the full factorial analysis, main effects were obtained for type of label, gender, and computer platform; there were no main effects for expertise or age, and no interaction effects. Given those results, the analysis was run again with only main effects in the model. These results are shown in Table 3. A Scheffé post-hoc procedure (alpha = .01) was used to examine the levels of the main effects. The three types of label were all significantly different from each other, with the standard label most likely to engender a reported increase in appreciation, followed by the relevance label and finally the QA/metaphor label. The effect sizes (Cohen's $d$) were as follows: standard to relevance, .244; standard to QA/metaphor, .492; relevance to QA/metaphor, .249. Females gave significantly higher ratings than did males (Cohen's $d$ = .211). For type of computer platform, those using a smart phone were more likely to report increased levels of appreciation as compared to the other types of platforms (Cohen's $d$) compared to laptop = .377, compared to desktop = .241, compared to tablet = .220).

Table 3

*ANOVA Main Effects for 3 (label type) * 2 (gender) * 3 (age groups) * 3 (expertise) * 4 (computer platforms) for Appreciation*

| Source* | n | M | SD | df | F | $\eta^2_p$ | p |
|---|---|---|---|---|---|---|---|
| Label Type | 1,626 | | | 2 | 34.78 | .041 | <.001 |
|     QA/Metaphor | 529 | 6.87 | 2.13 | | | | |
|     Relevance | 536 | 7.35 | 1.87 | | | | |
|     Standard | 561 | 7.82 | 1.74 | | | | |
| Gender | 1,626 | | | 1 | 12.26 | .008 | <.001 |
|     Male | 1,015 | 7.20 | 1.98 | | | | |
|     Female | 611 | 7.61 | 1.90 | | | | |
| Age Groups | 1,626 | | | 2 | 2.53 | | *ns* |
|     18-34 | 540 | 7.69 | 1.83 | | | | |
|     35-54 | 490 | 7.20 | 1.99 | | | | |
|     55 and above | 596 | 7.17 | 2.01 | | | | |
| Grouped Expertise | 1,626 | | | 2 | 0.28 | | *ns* |
|     1-3 | 443 | 7.40 | 2.01 | | | | |
|     4-6 | 637 | 7.36 | 1.90 | | | | |
|     7-10 | 546 | 7.30 | 1.98 | | | | |
| Computer Platform | 1,626 | | | 3 | 9.14 | .017 | <.001 |
|     Smart Phone | 485 | 7.81 | 1.84 | | | | |
|     Tablet | 115 | 7.39 | 1.88 | | | | |
|     Laptop | 474 | 7.09 | 2.02 | | | | |
|     Desktop | 552 | 7.35 | 1.96 | | | | |

*NOTE.* No interaction effects were obtained.

**Label Preferences**

In the analyses so far, participants were looking at a single label and rating it. For the last set of analyses, participants directly compared the three different label types for a single image (assigned at random). They were asked the following questions for the three labels:

1. Which provides the best explanation for the image? [label 1, label 2, or label 3]

2. Which explanation makes you want to learn more about this topic? [label 1, label 2, or label 3]

3. Which explanation increases your overall appreciation of the image best? [label 1, label 2, or label 3]"

We summed responses over the four labels to find the overall preferences. For the question 1 (best explanation), the percentage of preference was: Standard (52.8%), Relevance (31.9%), and Question/Metaphor (15.3%). These differences were tested with a goodness of fit chi square test and were significant, $X^2$ (2, N = 1730) = 365.79, $p$ < .001. Tests for each pair of percentages showed that all differences were significant at $p$ < .001. For question 2 (learn more), the percentage of preference was: Standard (31.6%), Relevance (50.9%), Question/Metaphor (17.5%). The differences were significant, $X^2$ (2, N = 1730) = 292.93, $p$ < .001. Again, all pairs were significantly different at $p$ < .001. For question 3 (overall appreciation), the percentage of preference was: Standard (48.8%), Relevance (35.6%), and Question/Metaphor (15.5%). The differences were significant, $X^2$ (2, N = 1730) = 291.79, $p$ < .001. All pairs were significantly different at $p$ < .001.

**Questions for An Astronomer**

After viewing and rating the assigned label for each image, participants were asked, "What question might you ask if an astronomer were with you now?"

**Cassiopeia A.** There were 1,389 questions for this image from $n$ = 1,151 of the participants. The leading question and metaphor label for this image concerned how exploded stars are similar to bees; the relevance label concerned the responsibility of supernovas in creating the oxygen we breathe, the iron in our blood, and the calcium in our bones. The standard label did not include the information from either of those labels. For the most part, the responses reflected the non-expert nature of the sample. Table 4 shows the categories of the responses, along with counts for each category and representative questions for each category. It is interesting to note the large number of questions that were somewhat technical in nature but are worded in a non-technical fashion.

Table 4

*Responses to the Open-Ended Item "What question might you ask if an astronomer were with you now?" for Cassiopeia A*

| Category | n | Representative Responses |
|---|---|---|
| Technical Question Non-expert | 391 | How powerful is the explosion during a supernova? (F, 4, QA/M) |
| | | How do new stars form after the event like supernova as there is no fuel left to carry on? M, 1, QA/M) |
| | | What elements are detected in this supernova? (F, 5, S) |
| Color/Color Assignment | 215 | What do the different colors represent?  (F, 8, R) |
| | | Are those colors in the photo true, or are they enhanced? (M, 5, QA/M) |
| | | What causes the supernova to be different colors? (F, 4, S) |
| Elements | 172 | Do comets carry these same elements? (M, 1, QA/M) |
| | | What specific conditions of a supernova make specific elements? Is it just a random or is there a pattern? (M, 7, R) |
| | | Do we know what elements came from this supernova? (F, 7, S) |

*(continued)*

| | | |
|---|---|---|
| Effect on Earth/Our Solar System | 153 | If our solar system's material is the result of a supernova explosion, where is our "parent" star now? (M, 8, S) |
| | | How many supernovas would be needed to provide the material found in planet Earth? (M, 3, R) |
| | | What nearby stars potentially could go supernova and are they dangerous to us? (F, 7, R) |
| Size/Distance From Earth | 123 | How much larger is the debris field now compared with the original size of the star? (F, 2, S) |
| | | How big is what we're seeing in this image compared to the size of our solar system or the Milky Way Galaxy? (M, 6, S) |
| | | How far away is Cassiopeia A from the Earth? (M, 5, R) |
| Time/Speed | 121 | How long does it take for a star to collapse in this way? (F, 5, R) |
| | | How is it still there if the supernova happened 300 years ago? How do you know it was 300 years? (M, 7, S) |
| | | Do we know the distance and speed of the traveling elements? (M, 4, QA/M) |
| Technical Question Showing Expertise | 74 | What are the heaviest elements synthesized in type 2 supernova compared to type 1a? (M, 10, QA/M) |
| | | Why didn't the superheated gas and dust become a plasma? (F, 8, S) |
| | | Are the elemental abundances in the outward facing bow shocks distinct from the rest of the object? (M, 10, S) |
| General Non-Expert Question | 69 | What would it look like if I was floating in space next to it? Well not right next to it. But close enough to identify it. (F, 3, S) |
| | | Can I see this if I look up in the sky? Or with a telescope? (M, 6, R) |
| | | How are you able to take this picture if it happened 300 years ago? (F, 6, S) |
| Metaphor (Bees) | 40 | Isn't it really more like a dandelion exploding and dispersing its seeds without carrier? (F, 1, QA/M) |
| | | The bee metaphor makes the idea of supernova more relatable. (M, 3, QA/M) |
| | | The allusion to bees and pollen could be taken by some people to imply a guided dispersion. (M, 1, QA/M) |

| Other Comments | 31 | How was it received by religious folk of the day? (F 7, S) |
| | | AHHHHHHHHHHHHH this is so awesome! (F, 2, R) |
| | | Probably the most amazing thing I've ever learned is that I am composed from supernova ejecta. (M, 3, QA/M) |

*NOTE.* For information in parentheses, M/F indicates gender, 1-10 indicates level of self-reported expertise, and label type is indicated by QA/M, R, and S for Leading Question/Metaphor, Relevance, and Standard, respectively.

**Solar Flare.** There were 1,386 questions for this image from $n = 1,176$ of the participants. The QA/ metaphor label for this image concerned whether storms on the Sun are like those here on Earth; the relevance label pertained to how storms on the Sun can affect disrupt orbiting satellites, affecting GPS and cell phone service, and allowing us to see auroras. The standard label referenced the aurora associated with the image. Although there was overlap with Cassiopeia A in terms of the categories of the questions, and again, for the most part, the responses reflected the non-expert nature of the sample, the numbers within categories differed. Table 5 shows the categories of the responses, along with counts for each category and representative questions for each category. It is interesting to note the large number of questions that addressed the effect of solar flares on the Earth and its inhabitants.

Table 5

*Responses to the Open-Ended Item "What question might you ask if an astronomer were with you now?" for Solar Flare*

| Category | n | Representative Responses |
|---|---|---|
| Effect on Earth/ Our Solar System | 382 | Could we harness this energy? (F, 3, QA/M) |
| | | Are some planets more susceptible to the sun's magnetic force? (M, 5, R) |
| | | Can a solar storm kill all life on Earth? (M, 8, R) |
| Technical Question Non-expert | 262 | Can other phenomena occur from these solar storms? (M, 2, QA/M) |
| | | Do these flares have a strong effect on Mercury and Venus, which are closer to the sun than the Earth is? (F, 4, R) |
| | | How do these solar flare events indicate what is going on inside the sun? (M, 3, R) |
| Frequency/ Duration/ Predictability | 166 | Can they be predicted in any way? How big are they? How long do they last? (M, 6, QA/M) |
| | | As the sun ages, should we expect more or fewer solar flares? (M, 7, QA/M) |
| | | How far in advance can solar storms be predicted? Has the number of solar storms increased or decreased over the years? (F, 3, R) |
| Cause of Solar Flares? | 127 | Can you give a detailed explanation of what causes CMEs? (M, 8, S) |
| | | Cause of solar storms? Distinction between storms, flares, sunspots, prominences, etc. (F, 7, QA/M) |
| | | Earth has storms because we have an atmosphere. What causes storms on the sun? (M, 4, S) |
| Size/Distance | 98 | How far do flares travel? (M, 5, R) |
| | | How far away from Earth are they before we can see them? (M, 8, QA/M) |
| | | How long is the flare? How big can they get? How big is it in comparison to Earth? (F, 6, S) |
| Particles/Elements/ Plasma | 95 | Are there different kind off particles and will they arrive in earth's atmosphere at different times? (F, 6, S) |
| | | Do elements form in the flares? Why do flares cause electromagnetic disturbances? Is it because they are plasma? (F, 3, R) |

*(continued)*

| | | |
|---|---|---|
| | | I'd like to know the plasma of what elements make up the flare? (1, 7, S) |
| North/South Poles and Auroras | 89 | Is aurora activity always connected to solar flares? (M, 2, S) |
| | | How do atmospheric conditions affect the extent of auroras-- or do they? Does temperature affect how far north or south auroras reach? (F, 8, QA/M) |
| | | How do solar flares create auroras? (F, 3, S) |
| Technical Question Showing Expertise | 71 | Why the temperature differences in the coronal ejection and the photosphere? (M, 8, QA/M) |
| | | Is the filament held together by a el-mg field? (M, 7, S) |
| | | Please explain to me the differences/correlations between Ångströms and nanometers in reference to wavelength. (M, 8, S) |
| Metaphor (Weather/Storms) | 67 | How can liquid fire become a storm? Does it rain fire like clouds rain? (M, 5, QA/M) |
| | | Do the solar storms affect climate change on Earth (and other planets)? (M, 9, QA/M) |
| | | Do all types of storms on the sun produce the plumes of superheated gas, or do some stay contained on the sun? (F, 2, QA/M) |
| Other Comments | 29 | This is an amazing picture. I shared it with all my friends and they agree. (M, 5, S) |
| | | That is one freaking awesome sun picture! (M, 3, QA/M) |
| | | That picture is just AWESOME! (F, 4, S) |

*NOTE.* For information in parentheses, M/F indicates gender, 1-10 indicates level of self-reported expertise, and label type is indicated by QA/M, R, and S for Leading Question/Metaphor, Relevance, and Standard, respectively.

**Sagittarius A\*.** There were 1,292 questions for this image from $n$ = 1,222 of the participants. The leading question and metaphor label for this image asked if black holes are like giant cosmic vacuum cleaners; the relevance label concerned how black holes are more responsible for the development of galaxies that surround them than any destruction. The standard label did not contain either of these pieces of information. As with the previous two images, there was overlap with the categories of the questions asked, and the responses indicated the non-expert nature of the sample. Table 6 presents the categories of the responses, along with counts for each category and representative questions for each category. It is interesting to note how many questions concerned where the black hole was in the picture, and that in place of comments expressing awe, there were a number of science fiction and time travel questions.

Table 6

*Responses to the Open-Ended Item "What question might you ask if an astronomer were with you now?" for Sagittarius A\**

| Category | n | Representative Responses |
| --- | --- | --- |
| Technical Question Non-expert | 349 | Are black holes common? Is there more than one per galaxy? Can they die? (M, 3, QA/M) |
| | | Can we send a probe to the black hole? (M, 5, S) |
| | | How close is "too close" to a black hole? (F, 4, R) |
| Where Is It? | 268 | Where exactly is the black hole in the image? (F, 3 QA/M) |
| | | Where exactly in this image is the black hole? In the exact center? In the bright blue high-energy shot just to the lower right of center? Can we actually SEE a black hole? (or rather, do we just see a big spot of lack-of-light?) (F, 7, S) |
| | | Where is the black hole in this picture? What do the colored "clouds" (red, amber, blue) signify and what is that bright blue body? (M, 3, R) |
| How Is It Created and/or How Does It Create Cosmic Structures? | 188 | What are the leading theories on the formation of supermassive black holes? (F, 8, S) |
| | | Which "cosmic structures" are black holes responsible for creating? (M, 6, R) |
| | | How does a black hole contribute to the growth of a galaxy? (M, 7, R) |

*(continued)*

| | | |
|---|---|---|
| Effect On Earth/<br>Our Solar System | 126 | What makes it supermassive and does it exhibit any effect on Earth? (M, 5, S) |
| | | What effect might a black hole have on the star forming areas of a galaxy? (F, 3, QA/M) |
| | | How might the future of our galaxy be effected [sic] by this object? (M, 5, R) |
| Size/Distance | 108 | Black holes are appearing and eventually growing but never disappearing. Are distances between them constant? (M, 3, S) |
| | | Is there a telescope on earth that can see the black hole? (F, 5, QA/M) |
| | | How big can black holes be? Can they consume galaxies? (F, 7, R) |
| Color/Color Assignment | 65 | What do the colored "clouds" (red, amber, blue) signify and what is that bright blue body? (M, 3, R) |
| | | I don't see much if any green (medium energy) in this image but do see orange and yellow. What do those colors represent? (M, 7, S) |
| | | What do the different colors represent and why are there regions that appear to be separated by color? What kind (visible, UV, radio…?) of light is this picture taken in? (F, 9, R) |
| Technical Question Showing Expertise | 63 | Does the temperature of the X-ray emitting gas trace the gravitational potential around Sgr A*? (M, 9, S) |
| | | I know black holes accrete matter and cause massive frictional heating that can power the galaxy but that seems impossible considering the distance between the black hole and other stars, and how exactly would emitted energy "power" the galaxy? (M, 7, R) |
| | | How much mass has to be ingested to make a black hole active? (F, 8, QA/M) |
| Metaphor (Vacuum Cleaner) | 47 | So if a black hole isn't a vacuum cleaner, what is it like instead? (M, 6, QA/M) |
| | | Black holes don't "suck"; try and use analogy to explain the gravity well. (M, 8, QA/M) |
| | | If we use our vacuum cleaner the dust is accumulated in the space given in the vacuum cleaner. Like the vacuum cleaner where will be the cosmic dust or stars go or be stored after black hole suction? (M, 9, QA/M) |
| Time/Speed | 42 | Why does the escape velocity of a black hole exceed the speed of light? (F, 8, QA/M) |
| | | To what extent does time dilation play a role as we approach the event horizon of these beautifully mystical objects? (M, 8, R) |
| | | Is time really affected by the gravitational pull of a super massive black hole? (M, 6, S) |
| Science Fiction/Movies | 36 | Any chance that black holes are portals to other parts of space, like worm holes? (M, 1, R) |
| | | What are the chances that black holes are the time travel portals portrayed in science fiction stories?? (F, 3, QA/M) |
| | | No one thinks black holes are like giant vacuum cleaners! Everyone knows the analogy is like flushing a toilet or using the garbage disposal in a sink - including the "backspray" of materials back into the cosmos. That's how so many sci-fi movies have the heroes escaping by being flung AWAY from the black holes while the bad guys get sucked in and dematerialized. (F, 3, QA/M) |

*NOTE.* For information in parentheses, M/F indicates gender, 1-10 indicates level of self-reported expertise, and label type is indicated by QA/M, R, and S for Leading Question/Metaphor, Relevance, and Standard, respectively.

**Pinwheel Galaxy (M101).** There were 1,264 questions for this image from *n* = 1,191 of the participants. The leading question and metaphor label for this image concerned how pinwheel galaxies are similar to water swirling down a drain; the relevance label centered on how this "sister" galaxy allows us to learn about our own. The standard label did not contain either of these pieces of information. Again, there was overlap with the categories of the questions asked, and the responses indicated the non-expert nature of the sample. Table 7 presents the categories of the responses, along with counts for each category and representative questions for each category. There was wide interest in how the Pinwheel Galaxy compared to other types of galaxies and our own Milky Way, along with questions about the spin and about time in relation to the image as captured.

Table 7

*Responses to the Open-Ended Item "What question might you ask if an astronomer were with you now?" for Pinwheel Galaxy (M101)*

| Category | n | Representative Responses |
|---|---|---|
| Technical Question Non-expert | 318 | Do you know if there are any supernovas in this galaxy? (F, 4, S) |
| | | Are there any factors that influence orientation of the accretion disk of a galaxy, or is it mostly random? (M, 6, R) |
| | | Are you able to discern whether this galaxy has a black hole at its core? If so, how much would this black hole contribute to the spin of the galaxy? (F, 3, QA/M) |
| Pinwheel and Other Galaxies | 196 | Are there a set number of spiral arms or do they vary among galaxies? (F, 6, QA/M) |
| | | Are there images of other pinwheel-shaped galaxies of different ages? I'd like to see older and younger ones side-by-side (F, 4, R) |
| | | What makes this galaxy spiral compared to another galaxy? Can galaxies switch between different forms, and if so, how? (M, 7, S) |
| Comparison to Earth/ The Milky Way | 183 | What did the Milky Way look like in the same time – 25 million years ago? (M, 3, S) |
| | | Are scientists able to ascertain the comparable size of the Pinwheel Galaxy to the Milky Way? (F, 2, R) |
| | | What do you think will occur to Earth when the Milky Way and Andromeda galaxies collide? (F, 7, QA/M) |
| Size/Distance across it? | 148 | How many stars are in this galaxy? What is the distance (M, 4, QA/M) |
| | | How accurate is the distance estimate to M101? (M, 8, S) |
| | | What is the mass of this galaxy? How far away is it? What accounts for the slight asymmetry? (F, 8, R) |
| Spin/Direction of Spin | 122 | Do pinwheel galaxies all turn in the same direction or can they turn in different directions? Or is the direction just a matter of the frame of reference that it's being viewed from? (M, 5, S) |
| | | Do galaxies spin the other way when viewed from south of the equator? (M, 7, QA/M) |
| | | How fast does a spiral galaxy spin? (F, 3, R) |
| Questions About Time | 107 | Does the shape of this galaxy show any space/time deformities caused by the black hole's rotation? (M, 7, QA/M) |
| | | Can we look further back in time of this structure? How might have it looked if we were able to view it during the Miocene? How might it look now? In the future? (M ,4, S) |
| | | Could other distant galaxies we see actually be ours from 26,000,000 years ago, just seen by a sort of cosmic lensing and we aren't aware? (F, 3, S) |
| Technical Question Showing Expertise | 88 | How many bands was this image taken in? Can you derive stellar populations data from it? (M, 10, S) |
| | | Which stellar population synthesis models have been typically assumed (for this galaxy) when observers estimate its physical properties from spectra? (F, 10, QA/M) |
| | | Does M101 contain any Wolf Rayet nebulae similar to ones found in the Milky Way and the LMC? (M, 7, QA/M) |
| Color/Color Assignment | 41 | What do the different colors represent? (M, 7, QA/M) |
| | | If the white areas are clusters of stars, what are the reddish streaks? (F, 6, R) |
| | | Lots of images of spiral galaxies show pink regions. Why don't we see any in this picture? (M, 10, S) |
| Metaphor (Water Swirling Down a Drain) | 24 | What's causing the galaxy to look like water going down the drain? What's causing the "low pressure"? (M, 5, QA/M) |
| | | If a galaxy is like water going down a drain, how come we can see so many spirals? Shouldn't the pattern get destroyed as the inner parts go around faster than the outer ones? (M, 4, QA/M) |
| | | Is there a black hole at the center of this galaxy? Is it a spiral because it IS literally draining into the black hole at the center of the galaxy? (F, 5, QA/M) |
| Other Comments | 37 | Thanks for being awesome! (M, 8, S) |
| | | While people might differ on the particulars, it would be a safe assumption to say that most everyone who would look at this galaxy would say it was beautiful. There is an aesthetic quality that captivates the imagination and takes our breath away. (M, 6, S) |
| | | Just a comment to say but how beautiful is that? (F, 3, R) |

*NOTE.* For information in parentheses, M/F indicates gender, 1-10 indicates level of self-reported expertise, and label type is indicated by QA/M, R, and S for Leading Question/Metaphor, Relevance, and Standard, respectively.

**Final Question.** There were 603 final comments. The majority were messages of thanks and appreciation or enjoyment with participating, such as, "Your photographs and explanations are a wonderful way to lift the interest if the layman about astrophysics. They are beautiful and there is so much to learn. Thank you for sharing your knowledge" (F, 4, QA/M). Others commented on having learned from participating in the survey and wanting to know how to participate in future studies: "This was really interesting, and it made me want to learn more about the topics covered!" (F, 1, R). "Taking this survey created a desire to learn more about the topics discussed. Thank you!" (M, 6, R). Still others had suggestions (e.g., "I'm 92; increase your age groups!") or critiques (e.g., "Use different colors. The ones used are garish."). And finally, over half of the comments pertained to the aesthetics of the images, using words like beautiful, gorgeous, awesome, spectacular, and aesthetically pleasing or appealing, and powerful.

## Discussion

This study extended previous research on public perceptions of the aesthetic appeal, educational value, and communication of underlying messages associated with astronomical images by examining the use of leading questions/metaphor and relevance to everyday life in labels. With regard to the results as they related to the research questions, first, in considering perceived knowledge of the objects in the labels, the relevance label led to an increase in perceived knowledge (Cohen's *d* = .299). The relevance label would be classified by Leder, Carbon, and Ripsas (2006) as an *elaborative label*. The standard label ("descriptive" in Leder et al.'s terms) only showed an effect size of .080 (*ns*). Thus, these two findings are consistent with the ideas put forward by Leder et al. However, the QA/metaphor label, which had been found useful in previous research (Gutwill, 2006; Jones, 1995; Litwak, 1996) and specifically with astronomical images (Smith, 2014; Smith, et al., 2011; Smith et al., 2014; Smith, Smith, Arcand, Smith, & Bookbinder, 2015), did not lead to increased perceived knowledge in this study ($d = -.081$, $ns$). Furthermore, when we made direct comparisons of labels, the Standard label was more popular than the Relevance label or the Question/Metaphor label in terms of providing the best explanation and for overall appreciation of the image. The Relevance label was the second preference for each of those questions, and led the preferences in terms of participants wanting to learn more. Thus, these findings only show partial support for previous work in this area.

Although the results did not confirm previous research on the type of label that engendered the most positive responses, what is clear from these results is that the notion of having informative labels accompanying astronomical images is crucial to their understanding and appreciation. This is consistent with the research on works of art and labels (Belfiore & Bennett, 2007; Kreitler & Kreitler, 1972).

In terms of wanting to learn more, relevance was the preferred type of label. Those who reported low levels of expertise were less likely to want to learn more as compared to those with mid- or high levels of self-reported expertise. When type of computer platform was considered, participants who took the survey using a smart phone were more likely to indicate that they wanted to learn more than those on laptops, desktops, or tablets. The findings for relevance and level of expertise are relatively straightforward to interpret.

Learning something interesting and personally relevant about an image might well lead to the participant wanting to know more. Similarly, those who have taken up astronomy to some degree (over people with low levels of expertise) have already indicated a desire to know about the topic.

The finding that people on a smart phone were more likely to say they wanted to learn more (than people on a tablet or computer) is a bit perplexing. It is possible that someone who is engaged with a study on astronomical images and is using a smaller device might simply be inherently more interested in astronomy and eager to participate in a study than others on other types of devices (since they are willing to do so on a device that provides a rather small image). This would be something to explore in a future study.

When examining aesthetic appreciation for deep space images, however, it is interesting to note that the original label was rated highest overall, followed by the relevance label, and then the leading question/metaphor label. Here, the results appear to stand in direct contradiction to Leder et al. (2006) and the other research on labels as well. Once again, those using smart phones gave higher ratings overall than those using other computer platforms; females gave higher ratings of appreciation than did males. What is perhaps not astounding but rather is reassuring is that level of self-reported expertise did not prove to be significant, perhaps indicating that no matter your background in astronomy or type of label, the "wow factor" is very real. A potential explanation for the preference for the original label is that it might encourage more focus on the image itself, as compared to the QA/metaphor or relevance labels. With the additional information or context provided by the QA/metaphor and relevance labels, attention might be pulled away toward thinking about the ideas presented in the label rather than being placed with image. Thus, the participant is less engaged with the "awe factor" of the image and more engaged with the concepts being presented.

This is not to say that the images do not engender appreciation, as evidenced by the overall high mean levels on the question, as well as the responses to the final item that requested additional comments. Here, appreciation was apparent in abundance from participants across demographic groups.

The responses to the open-ended items requesting questions for an astronomer not only indicated that the participants represented a wide range of expertise, but also underscored the differences among those levels. For the less expert, concern for the effects of events in space on the Earth, our solar system, or humans in general was evident, as well as the need for providing everyday anchors for the underlying science. Those with less expertise wanted to know about the images, and welcomed learning interesting and new facts, especially as they applied on a personal level. It is also evident that understanding is difficult, or even misleading, unless information about scale, location, distance, accessibility for viewing, or regularity or predictability of events in space is provided. There is some indication that it might even be useful to exploit popular culture, especially sci-fi movies and books, to appeal to and communicate with the public. For expert viewers, perhaps a set of more technical labels is required that acknowledges background in the science underlying the images.

The conclusion and main take-away message from this study is that one size in labels is not going to fit all; different messages are needed for different audiences. The findings indicate that we've only begun to explore ways to communicate information from astronomical images with our audiences. What should follow is research that examines how to reach viewers based on individual characteristics, certainly taking into consideration background or expertise, but also in terms of cognitive, aesthetic, emotional, recreational, and educational needs or wishes. Further attention is also needed to the types of computer platform that are used to view images, as compared to images in a museum setting.

As with any research, there are limitations to this study that caution against generalization beyond the sample and procedures described. A specific issue with the design of this study concerns a potential confound with the metaphors and the relevance to everyday life used in the labels, in particular for the images of the Sun and Sagittarius A*. The descriptions of storms on the Sun, auroras, black holes gobbling things, and vacuum cleaners may not have been sufficiently different to warrant firm conclusions based on the data collected. A future study might invite focus groups to brainstorm ideas for labels that would provide metaphors and relevance for images from space, and then investigate the efficacy of those ideas for communicating to and engaging the public.

Where next? Images from space have the potential to provide aesthetic enjoyment, educate, and convey powerful messages. This study provides a wide foundation for additional research in the service of increasing those, especially to the lay public and those who are not experts in the field of astronomy.

To borrow a phrase, next stop is "Infinity and Beyond."

**Acknowledgements**

Partial support for this project came from NASA's Chandra X-ray Observatory, operated by the Smithsonian Astrophysical Observatory (SAO) under NASA Contract NAS8-03060. The authors gratefully thank Shelley Morgan from the University of Otago. The authors also thank Kristin Divona and Megan Watzke from the Smithsonian Astrophysical Observatory. Finally, the authors acknowledge Dr. Robert Nemiroff, Dr. Jerry Bonnell and APOD (Astronomy Picture of the Day) for helping us promote this research survey.

Black Holes and Vacuum Cleaners 45

**Appendix**
**Image Labels**

*Cassiopeia A*
Leading question/Metaphor:  How are exploded stars similar to bees? When stars of a certain size run out of fuel, they collapse onto themselves and explode. These events, known as supernovas, hurl the remains of the stars -- which can contain elements such as oxygen, iron, and sulfur to name a few -- out into space. This process disperses these important elements into the material that will eventually form future generations of stars and planets. Similar to how bees spread pollen by traveling from one flower to another, supernovas help seed key cosmic ingredients across space. (93 words)

Relevance:  When stars of certain size run out of fuel to burn, they collapse onto themselves and explode. These events, known as supernovas, hurl the remains of the stars -- which can contain elements such as oxygen, iron, and sulfur -- out into space. In fact, the only place that scientists know such vital elements for life can be made is in the nuclear furnaces of stars. To put it another way, the oxygen we breathe, the iron in our blood, and the calcium in our bones was originally created inside distant stars and distributed to our stellar and planetary ancestors through supernovas. (102 words)

Standard: Cassiopeia A is a young supernova remnant in our Milky Way Galaxy, believed to be the leftovers of a massive star that exploded over 300 years ago. The material ejected during the supernova smashed into the surrounding gas and dust at about 16 million kilometers per hour. This collision superheated the debris field to millions of degrees, causing it to glow brightly in X-rays as seen here by NASA's Chandra X-ray Observatory. (72 words)

*Solar Flare*
Leading question/Metaphor:  Are storms on the Sun like those here on Earth? Storms on Earth can take many forms, such as snow, rain, and wind. The Sun also experiences storms, but they look different from the terrestrial variety. Storms on the Sun result in giant plumes of superheated gas that travel at millions of miles per hour into the Solar System. When the particles from these storms hit the magnetic field around the Earth, they can stream down toward our planet's poles. This excites molecules in our atmosphere and can cause the beautiful sights of auroras (the "Northern or Southern Lights"). (99 words)

Relevance:  When storms erupt on the Sun, they can send massive amounts of energy and particles hurtling through space. If these solar storms are aimed toward Earth, they can disrupt orbiting satellites, affecting everything from GPS to cell phone service here on the surface. The output from these storms also can slam into our planet's magnetic field, sending energy and particles flowing down the magnetic field lines towards the Earth's poles. This influx of energy excites molecules in the Earth's atmosphere, generating a gallery of glowing colors in the sky. In the Northern Hemisphere, this phenomenon is known as the "Northern Lights," or auroras. (103 words)

Standard: On August 31, 2012, a long filament of solar material that had been hovering in the sun's atmosphere, the corona, erupted out into space at 4:36 p. m. EDT. The coronal mass ejection, or CME, traveled at over 900 miles per second. The CME did not travel directly toward Earth, but did connect with Earth's magnetic environment, or magnetosphere, causing an aurora to appear on the night of Monday, September 3. Pictured here is a lightened blended version of the 304 and 171 angstrom wavelengths taken from the Solar Dynamics Observatory. (89 words)

*Sagittarius A\**
Leading question/Metaphor: Are black holes like giant cosmic vacuum cleaners?
Black holes are often cast as the ultimate vacuum cleaners, sucking up everything that comes close to them. It is true that some material passes the "event horizon," that is, the point of no return, and will forever be lost within the black hole's gravitational grasp. However, the event horizon is a relatively tiny region around the black hole. Instead, black holes have a bigger influence on the material around them through their powerful jets and flares that send material *away* from the black hole and back into the cosmic environment. (99 words)

Relevance: Black holes have the reputation of gobbling up anything and everything that ventures within their reach. Although they do consume matter that comes too close to them, there are other ways that black holes play much larger and perhaps more important roles in the cosmic environment. For example, the giant black holes at the centers of galaxies are thought to be directly connected to the growth and development of the galaxies themselves. Astronomers are currently working to learn more about this symbiotic relationship, which has revealed that black holes are responsible for the *creation* of cosmic structures even more than their destruction. (102 words)

Standard: The supermassive black hole at the center of the Milky Way, known as Sagittarius A* (Sgr A*, for short), is over million times the mass of the Sun and is located about 26,000 light years from Earth. Astronomers use telescopes including the Chandra X-ray Observatory to monitor the behavior of Sgr A*. This Chandra image shows the region around the black hole in low, medium, and high-energy X-rays (red, green, and blue respectively.) Over the years, researchers have seen flares in X-ray light and witnessed other behavior to help explain the role that black holes like Sgr A* play in their host galaxies. (103 words)

*Pinwheel Galaxy (M101)*
Leading question/Metaphor: What does this galaxy have in common with water going down a drain?
The spiral shape of this galaxy immediately brings to mind the action of rotation. Like water winding its way down the drain in a sink or moist air spiraling its way into the low-pressure center of a hurricane, the rotation of a galaxy imprints its structure in the form of dense spiral arms that trace regions of star formation. This galaxy is commonly known as the Pinwheel Galaxy. Like our own Milky Way, it is classified as a spiral galaxy due to its swirling arms of stars, gas, and dust. (90 words)

Relevance: Like our own Milky Way, the so-called Pinwheel Galaxy is a spiral galaxy with spectacular arms of stars and dust. Because Earth is embedded within the Milky Way, we can never get a full perspective of how our own home galaxy looks or functions. Therefore, astronomers look outward into space to find comparable galaxies that can tell us more about our own galactic abode. With its face-on orientation to us, the Pinwheel Galaxy (formally known as Messier 101) allows astronomers the opportunity to examine the structure and behavior of this sister spiral galaxy to gain insight into our own galactic home. (101 words)

Standard: This image of M101 is the largest and most detailed photo of a spiral galaxy that has ever been released by the Hubble Space Telescope. M101, also nicknamed the Pinwheel Galaxy, lies in the northern circumpolar constellation, Ursa Major, at a distance of 25 million light-years from Earth. Therefore, we are seeing the galaxy as it looked 25 million years ago — when the light we're receiving from it now was emitted by its stars — at the beginning of Earth's Miocene Period, when mammals flourished and the Mastodon first appeared on Earth. The galaxy fills a region in the sky equal to one-fifth the area of the full Moon. (108 words)